\documentclass[]{aa}

\usepackage{graphics}
\usepackage{txfonts}

\newcommand{\ch}{} 

\begin{document}

\title{Electron-impact excitation of neutral oxygen
} 

\author{P. S. Barklem
}
 
\offprints{P. S. Barklem,
\email{barklem@astro.uu.se}}

\institute{Department of Astronomy and Space Physics, Uppsala University, Box 515, S 751-20 Uppsala, Sweden 
}

\date{Received (date) / Accepted (date)}

\abstract
{}
{To calculate transition rates from ground and excited states in neutral oxygen atoms due to electron collisions for non-LTE modelling of oxygen in late-type stellar atmospheres, thus enabling reliable interpretation of oxygen lines in stellar spectra.}
{A 38-state $R$-matrix calculation in $LS$-coupling has been performed.  Basis orbitals from the literature (Thomas~et~al.) are adopted, and a large set of configurations are included to obtain good representations of the target wavefunctions.  Rate coefficients are calculated by averaging over a Maxwellian velocity distribution.  }
{Estimates for the cross sections and rate coefficients are presented for transitions between the seven lowest $LS$ states of neutral oxygen.   The cross sections for excitation from the ground state compare well with existing experimental and recent theoretical results.   }
{}  


\maketitle

\section{Introduction}
\label{sect:intro}

Oxygen is one of the most abundant and important elements in the universe.  Oxygen is widely accepted to be predominantly produced in core-collapse supernovae resulting from massive stars, and thus detailed study of the chemical evolution of oxygen allows us to understand star formation history and the initial mass function, as well as supernovae yields.  The main observational evidence regarding the evolution of oxygen in the Galaxy comes from the measurement of oxygen abundances in the atmospheres of cool stars.  The magnitude and trend of measured oxygen abundance with age or metallicity is not yet well constrained, in particular the results at low metallicity differ between different studies (e.g. Israelian~et~al.~\cite{israelian98}, Boesgaard~et~al.~\cite{boesgaard99}, Nissen et~al.~\cite{nissen02}, Garcia-Perez et~al.~\cite{garciaperez06}).  The use of different spectral diagnostics is most likely the reason for these discrepancies.

The use of the \ion{O}{i} triplet at 777~nm is appealling as the lines are strong and unblended, even in metal-poor stars.  As they are in the visual region they are easily accessible with standard telescopes and spectrographs.  On the other hand, it is well known that these lines are not formed in conditions of local thermodynamic equilibrium (LTE) in cool star atmospheres (see Kiselman~\cite{kiselman01} and references therein).  If the non-LTE formation of these lines could be understood, then they would be very useful diagnostics for measuring stellar oxygen abundances.  However, we also note that the lines originate from highly excited levels, and are thus quite temperature sensitive, and furthermore may be saturated at higher abundances. 

The ability to solve the non-LTE problem is dependent on the availability of adequate atomic data for the radiative and collisional processes occuring to the oxygen atoms in the atmosphere.  The data for the collisional processes is the greatest stumbling block to reliable non-LTE modelling.  Collisions with electrons are particularly important, due to their large frequency (due to high velocity) and efficiency (collisions are not adiabatic).  Though collisions with hydrogen atoms are less frequent and expected to be less efficient (adiabatic) they may also be important due to sheer weight of number (typically $N_\mathrm{H} \ga 10^4 N_e$).  Though Kiselman~(\cite{kiselman01}) identifies the collisional cross-sections due to hydrogen as the most pressing issue since there is very little information on these processes (but see recent paper by Krems~et~al.~\cite{krems06}), the situation for electrons is still far from ideal.  Past studies of electron-impact excitation of neutral oxygen have predominantly focussed on excitation from the ground {\ch configuration}.  However, for the purposes of the non-LTE problem in stellar atmospheres, data for excited states are also needed.  Previous work on the oxygen non-LTE problem have used either approximate general formulae (e.g.\ Allende~Prieto~et~al.~\cite{allende03}) or a range of data calculated by different methods (e.g.\ Carlsson \& Judge~\cite{carlsson93}, Kiselman~\cite{kiselman93}).  In some cases data (except the simple formulae) do not exist, such as for the coupling between $LS$ states of different spin, e.g. triplet--quintet system coupling.  Kiselman performed test calculations for  three stellar atmosphere models and found that intersystem collisional coupling was probably not important.  Recent work by Fabbian~et~al.~(private communication) confirms the results of Kiselman for the considered models, but finds that triplet-quintet intersystem coupling may be important in the case of metal-poor turnoff stars (e.g. $T_\mathrm{eff}=6500$~K, $\log g=4$, $\mathrm{[Fe/H]}=-3$).

In any case, it is clear that a consistently computed set of reliable data for electron-impact excitation of oxygen is highly desirable for non-LTE studies of oxygen line formation in stellar atmospheres.  In this paper, we use the $R$-matrix approach to calculate electron-impact excitation cross sections for low energy collisions for transitions between the seven lowest $LS$ states of neutral oxygen, from which we calculate rate coefficients for cool star atmosphere temperatures.  We compare our results with experiment and those in literature, and discuss the differences. 

\section{Calculations}

The electron scattering cross sections are calculated using the $R$-matrix method (Burke et~al.~\cite{burke71}) in $LS$-coupling with readily available computer codes.  The internal region problem is solved using the the \texttt{RMATRX1} code (Berrington et~al.~\cite{berrington95}), and the external region scattering problem is solved using the  \texttt{FARM} code (Burke \& Noble~\cite{burke95}).  These papers, and references therein, should be consulted for detailed descriptions of the theory and the codes.  We now describe the calculations including the choice of orbital functions, basis configurations, target states and calculation parameters.

We employed the set of radial orbital functions from Thomas et~al.~(\cite{thomas97}). This basis consists of 1s, 2s and 2p orbitals from Clementi \& Roetti~(\cite{clementi74}), and 3s, 4s, 3p, 4p, 3d and 4d spectroscopic orbitals calculated by Thomas et~al.; this paper may be consulted for details.
Their basis also included pseudo-orbitals labelled $\overline{\mathrm{5s}}$, $\overline{\mathrm{5p}}$ and $\overline{\mathrm{5d}}$; however, these orbitals are not employed in our final calculations, as will be discussed below. 

A well-documented problem (e.g. Plummer et~al.~\cite{plummer04}, Berrington~et~al.~\cite{berrington88}) in $R$-matrix calculations for moderately complex atoms such as oxygen, is the balance between a good description of the target wavefunctions, and the problem of pseudo-resonances arising due to an inconsistent description of the $N$-electron target and ($N+1$)-electron problem.  Computer memory limitations often lead to use of inconsistent sets of $N$-electron and ($N+1$)-electron configurations.  {\ch We note the existence of methods which may allow this problem to be avoided or its effects drastically reduced, in particular the method developed by Gorczyca et~al.~(\cite{gorczyca95}) and the $B$-spline $R$-matrix approach with non-orthogonal orbitals (e.g. Zatsarinny \& Froese Fischer~\cite{zatsarinny00}, Zatsarinny \& Tayal~\cite{zatsarinny01}) a code for which was recently published by Zatsarinny~(\cite{zatsarinny06a})}.  With this in mind, we chose consistent $N$- and ($N+1$)-electron configurations for our calculations.  We included $N$-electron configurations arising from single and double excitations from  2s$^2$2p$^4$, but with maximum occupations of orbitals for different states as given in Table~\ref{tab:maxocc}.  For the ($N+1$)-electron configurations we account for all configurations arising from single, double and triple excitations from 2s$^2$2p$^5$ with the same maximum occupation numbers as for the target states.

As will be seen in the next section, our cross section results are significantly different from those of Thomas~et~al., in particular we do not find the same slow rate of increase of the cross sections near the threshold. Thomas et~al.\ chose configurations based on their expansion coeffients and the independent selection of particular configurations in the target and ($N+1$)-electron wavefunctions could have led to such pseudo resonances.  To test this, we also performed calculations selecting $N$-electron configurations in the same manner as Thomas et~al., i.e. based on their expansion coeffients, noting that the pseudo-orbitals $\overline{\mathrm{5s}}$, $\overline{\mathrm{5p}}$ and $\overline{\mathrm{5d}}$ were included in this calculation.  We obtained cross section results of similar form to those of Thomas~et~al., in particular we reproduced the slow rate of increase of the cross section near the threshold.  An additional test calculation was performed in which the pseudo-orbitals were not used but the same configurations, except those involving the pseudo-orbitals, were selected.  In this case the results were similar to our calculations where consistent $N$- and $(N+1)$-electron configurations were employed.  This suggests the inclusion of the pseudo-orbitals may be the main cause of the slow increase near threshold; however, one must note that the degree of incompleteness between $N$- and $(N+1)$-electron configurations is reduced due to the smaller set of basis orbitals. To resolve this definitively would require a calculation including the pseudo-orbitals and a large set of consistent $N$- and $(N+1)$-electron configurations, and this is presently beyond the computer resources available to us.  Based on these considerations and results, we decided to limit our orbital basis to the spectroscopic orbitals, enabling consistent sets of $N$- and $(N+1)$-electron configuration to be chosen.

\begin{table}
\tabcolsep 1.5mm
\begin{center}
\caption{Limits on occupation numbers for orbitals.}
\label{tab:maxocc}
\begin{tabular}{lrrrrrrr}
\hline
\hline
States                  & 3s & 3p & 3d & 4s & 4p & 4d & 4f  \\
\hline
Max.\ Occupation Number   &  2 &  2 &  2 &  2 &  2 &  2 & 1  \\
\hline
\end{tabular}
\end{center}
\end{table}

In the calculations we have included 38 target states in the close-coupling expansion which are listed in Table~\ref{tab:states}.    The model includes 19 spectroscopic target states, and 19 additional eigenstates which were included to partially account for coupling to the continuum.  Of these additional eigenstates, 14 correspond to observed autoionising states, while an additional 5 (pseudo-) eigenstates are included so that at least one state of all symmetries which dipole couple to the lowest seven target states are included.  We note that the inclusion of the autoionising and the pseudo-eigenstates {\ch does} not affect the cross sections greatly, and in particular the cross sections near threshold are basically unaffected.  However, their inclusion lessens the magnitude of pseudo-resonances occuring above the ionisation threshold.  Target energies for the 19 spectroscopic states are adjusted to the observed values in the scattering calculations.  Calculations were also performed with reduced numbers of target states.  A calculation including only the 19 spectroscopic states, and a calculation including the 33 physical states (spectroscopic and autoionising) were performed which we find useful for comparison below. 

The $R$-matrix calculations are performed with 50 continuum orbitals for each of channel angular momentum of $0\le l \le 25$. The $R$-matrix boundary is set at $a=56.4$~$a_0$.  Partial waves with total angular momentum up to and including $L=20$ are included, with full exchange included up to $L=12$. Test calculations indicate that all cross sections are well converged at energies below 40~eV even with only $L\le10$ included.

\begin{table}
\begin{center}
\caption{The included target states and their calculated theoretical excitation energies and observed energies; the difference is also shown. The observed energies are taken from NIST (see text).}
\label{tab:states}
\begin{tabular}{rrrr}
\hline
\hline
State               & Energy   & Energy     & Energy difference     \\
                    & (theory) & (observed) & (theory $-$ observed)\\
                    & [Ryd]    & [Ryd]      & [Ryd]      \\
\hline
\multicolumn{4}{c}{\underline{spectroscopic states}} \\
2p$^4$    $^3$P     & 0.000   & 0.000 & 	$ 0.000$ \\
2p$^4$    $^1$D     & 0.154   & 0.145 & 	$+0.009$ \\
2p$^4$    $^1$S     & 0.301   & 0.308 & 	$-0.007$ \\
2p$^3$3s  $^5$S$^o$ & 0.679   & 0.672 & 	$+0.007$ \\
2p$^3$3s  $^3$S$^o$ & 0.705   & 0.700 & 	$+0.005$ \\
2p$^3$3p  $^5$P     & 0.795   & 0.789 & 	$+0.006$ \\
2p$^3$3p  $^3$P     & 0.820   & 0.808 & 	$+0.012$ \\
2p$^3$4s  $^5$S$^o$ & 0.875   & 0.870 & 	$+0.005$ \\
2p$^3$4s  $^3$S$^o$ & 0.882   & 0.877 & 	$+0.005$ \\
2p$^3$3d  $^5$D$^o$ & 0.893   & 0.888 & 	$+0.005$ \\
2p$^3$3d  $^3$D$^o$ & 0.893   & 0.888 & 	$+0.005$ \\
2p$^3$4p  $^5$P     & 0.908   & 0.903 & 	$+0.005$ \\
2p$^3$4p  $^3$P     & 0.919   & 0.908 & 	$+0.011$ \\
2p$^3$3s  $^3$D$^o$ & 0.933   & 0.922 & 	$+0.011$ \\
2p$^3$3s  $^1$D$^o$ & 0.947   & 0.936 & 	$+0.011$ \\
2p$^3$4d  $^5$D$^o$ & 0.942   & 0.937 & 	$+0.005$ \\
2p$^3$4d  $^3$D$^o$ & 0.943   & 0.938 & 	$+0.005$ \\
2p$^3$4f  $^5$F     & 0.943   & 0.938 & 	$+0.005$ \\
2p$^3$4f  $^3$F     & 0.943   & 0.938 & 	$+0.005$ \\
\multicolumn{4}{c}{\underline{autoionising states}} \\
2p$^3$3s  $^3$P$^o$ & 1.033   & 1.039 & 	$-0.006$ \\
2p$^3$3s  $^1$P$^o$ & 1.047   & 1.057 & 	$-0.010$ \\
2p$^3$4s  $^3$D$^o$ & 1.135   & 1.116 & 	$+0.019$ \\
2p$^3$4s  $^1$D$^o$ & 1.137   & 1.120 & 	$+0.017$ \\
2p$^3$3d  $^3$P$^o$ & 1.149   & 1.125 & 	$+0.024$ \\
2p$^3$3d  $^3$D$^o$ & 1.150   & 1.133 & 	$+0.017$ \\
2p$^3$3d  $^1$P$^o$ & 1.150   & 1.133 & 	$+0.017$ \\
2p$^3$3d  $^1$D$^o$ & 1.151   & 1.133 & 	$+0.018$ \\
2p$^3$3d  $^1$F$^o$ & 1.151   & 1.134 & 	$+0.017$ \\
2p$^3$3d  $^3$S$^o$ & 1.150   & 1.134 & 	$+0.016$ \\
2s 2p$^5$ $^3$P$^o$ & 1.197   & 1.152 & 	$+0.045$ \\
2p$^3$4d  $^1$P$^o$ & 1.199   & 1.182 & 	$+0.017$ \\
2p$^3$4d  $^1$F$^o$ & 1.200   & 1.183 & 	$+0.017$ \\
2p$^3$4d  $^3$P$^o$ & 1.220   & 1.185 & 	$+0.035$ \\
\multicolumn{4}{c}{\underline{pseudo-eigenstates}} \\
          $^3$P     & 1.086   &       & 	         \\
          $^5$P     & 1.797   &       &                  \\
          $^5$P$^o$ & 1.898   &       & 	         \\
          $^5$D$^o$ & 1.911   &       & 	         \\
          $^5$S$^o$ & 1.930   &       & 	         \\
\hline
\end{tabular}
\end{center}
\end{table}

The accuracy of the wavefunctions may be tested, to some degree, by comparison with experimental observables such as excitation energies and oscillator strengths.  In Table~\ref{tab:states} the theoretical excitation energies are compared with experimental values taken from the NIST Atomic Spectra Database\footnote{National Institute of Standards and Technology, Atomic Spectra Database, http://physics.nist.gov.}.  The results agree quite well with experiment, particularly for the lower lying states which are of interest here.  In Table~\ref{tab:fvalues} the line strengths for transitions between low-lying states are compared with those from the NIST Atomic Spectra Database and Biemont~et~al.~(\cite{biemont91}).  Again, the agreement is satisfactory, indicating that the wavefunctions are of acceptable quality.

A correct treatment of long range dipole polarisation may be important for low energy electron-atom scattering calculations.  In the case of ions the long-range interaction is dominated by Coulomb forces, while for neutral atoms the lowest order interaction arises from the dipole moment induced in the atom by the electron's electric field.  Accurate representation of the dipole polarisation requires that contributions from all coupled channels, including the continuum, be accounted for.  In calculations, this may be achieved by the introduction of pseudostates, which mimic the contribution of a large number of channels (Damburg \& Karule~\cite{damburg67}).  A correct representation of the dipole polarisability is very important for elastic scattering calculations, particularly the differential cross sections (e.g. Plummer et~al.~\cite{plummer04}, Zatsarinny~et~al.~\cite{zatsarinny06}).  However, at least for oxygen, the available evidence points to this being less important in inelastic scattering processes.  Tayal~(\cite{tayal02}) found that for the $^3P$--3s$^3S^o$ and  $^3P$--3s$^3D^o$ transitions, inclusion of coupling to the continuum reduces the cross sections by 5 to 15\% and 5 to 27\% respectively.  Zatsarinny \& Tayal~(\cite{zatsarinny01}) and Plummer et~al.~(\cite{plummer04}) found small effects ($< 10$\%) for the $^3P$--$^1D$, $^3P$--$^1S$ and $^1S$--$^1D$ transitions.  Tayal~(\cite{tayal04}) concludes that the effect of coupling to the continuum is greater on resonance transitions than on forbidden transitions.  The effect is also expected to be mostly important above the ionisation threshold (e.g. Tayal~\cite{tayal02}), the differences seeming to be largest near the maximum in the cross section, but negligible within a few eV of the threshold.  

In our calculations we have not included any polarisation pseudostates, and only a fraction of the true dipole polarisability for each state is included.  For example, for the ground state the experimental value for the static dipole polarisability is found to be $\alpha=5.4\pm0.4$ a.u.\ (Alpher \& White~\cite{alpher59}), while theoretical values are typically between 4.6--5.4 a.u.\ (see results collected in Thomas et~al.~\cite{thomas97}).  We calculate for our model $\alpha=1.68$, and thus only about a third of the polarisability of this state is included.  Based on the above, noting that the majority of the transitions are forbidden, we expect that this may lead to an overestimate of cross sections for allowed transitions by of order 10\%.  Given that we are interested in quite low temperatures and thus low collision energies, the cross sections within a few eV of the threshold will dominate in the rate coefficients and thus the errors would most likely be even smaller.

\begin{table}
\begin{center}
\caption{Comparison of computed line strengths with data from the NIST compilation, and where available the theoretical calculation of Biemont~et~al.~(\cite{biemont91}).  }
\label{tab:fvalues}
\begin{tabular}{rrrr}
\hline
\hline
Transition                                   & $S$        & $S$     & $S$   \\
                                             &(this work) & (NIST)  & (Biemont~et~al.)      \\
\hline
2p$^4$   $^3$P      ---  2p$^3$3s $^3$S$^o$  &  2.211     & 2.001   &       \\
2p$^3$3s $^5$S$^o$  ---  2p$^3$3p $^5$P      &  130.8     & 128.4   & 123.7 \\
2p$^3$3s $^3$S$^o$  ---  2p$^3$3p $^3$P      &  81.8      & 86.3    & 89.5  \\  
\hline
\end{tabular}
\end{center}
\end{table}

\section{Results and Discussion}

\begin{figure*}
\begin{center}
\resizebox{170mm}{!}{\rotatebox{0}{\includegraphics{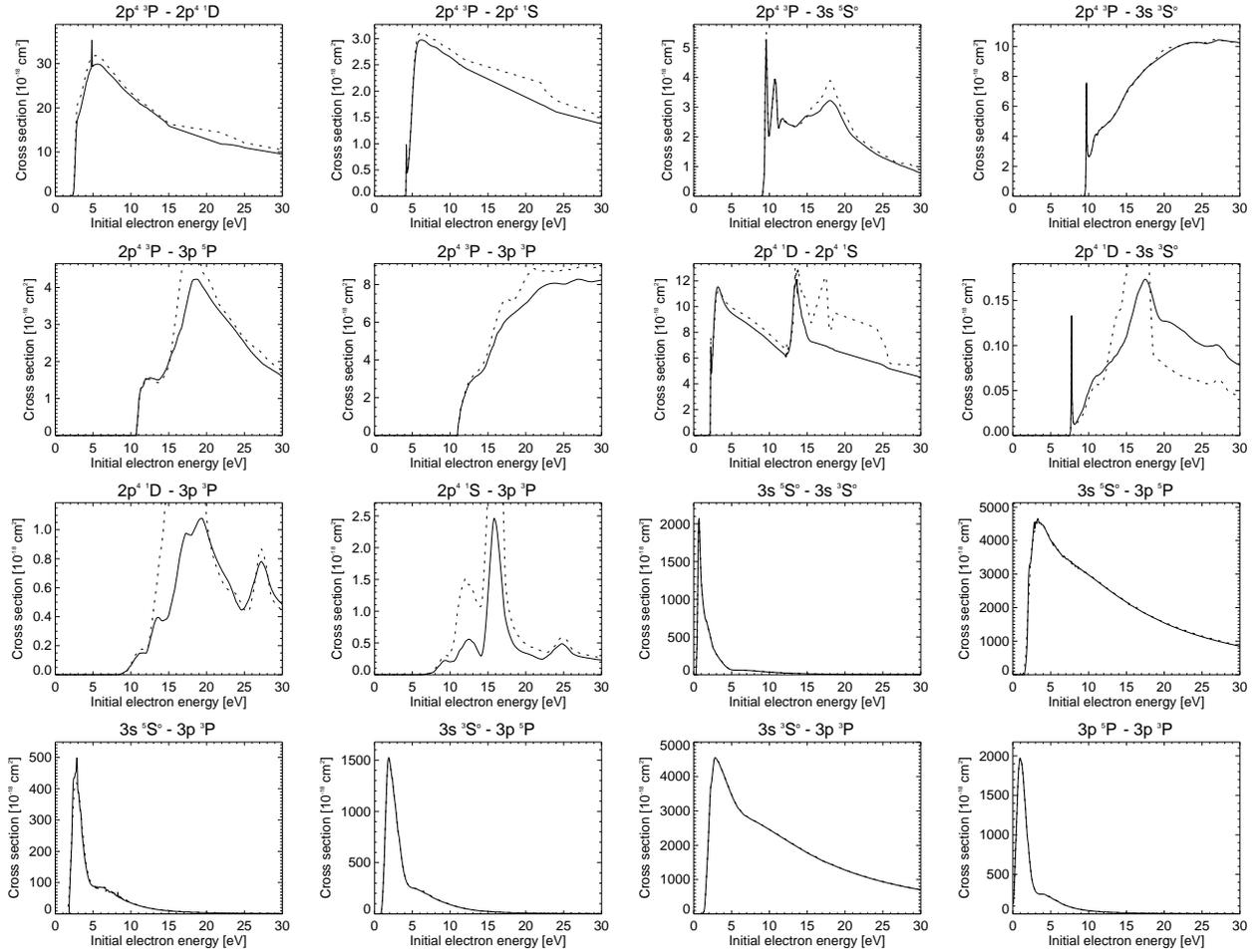}}}
\end{center}
\caption{The cross sections for electron-impact excitation as a function of initial electron energy between the seven lowest target states.  The full lines are the 38-state calculation, while the dotted lines are the 19-state calculation.  The cross sections are in units of $10^{-18}$~cm$^2$.}
\label{fig:cross}
\end{figure*}

\begin{figure*}
\begin{center}
\resizebox{\hsize}{!}{\rotatebox{0}{\includegraphics{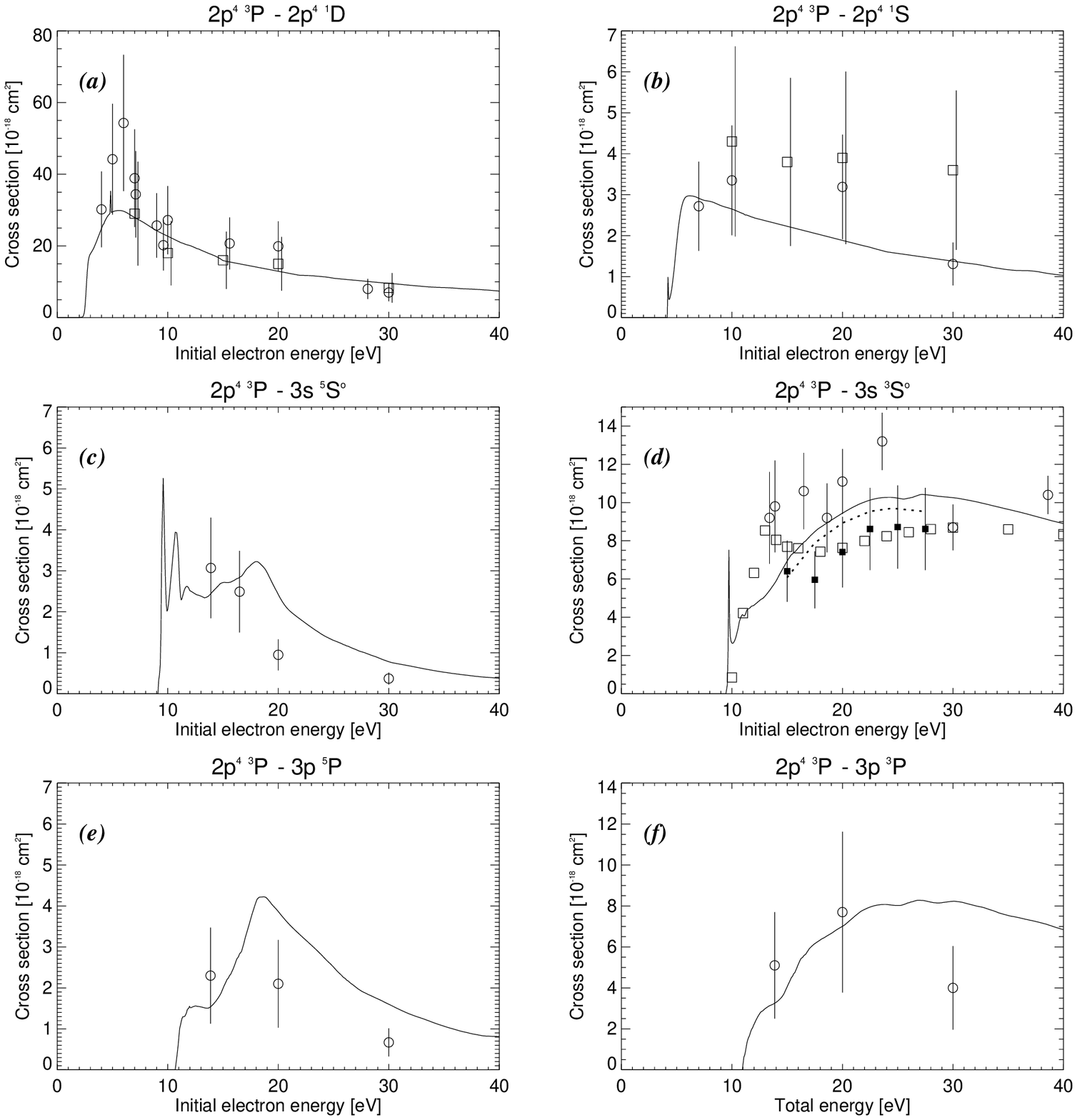}}}
\end{center}
\caption{The collision cross sections as a function of initial electron energy compared with experiment. The full lines are the results of the present work.  Experimental and other theoretical results in each case are from:
{\bf (a)} Doering~(\cite{doering92}), circles; Shyn \& Sharp~(\cite{shyn86a}, squares; in the latter case, the error bars are slightly offset for clarity. 
{\bf (b)} Doering \& Gulcicek~(\cite{doering89a}), circles; Shyn~et~al.~(\cite{shyn86a}), squares; in the latter case the error bars are slightly offset for clarity. 
{\bf (c)} Doering \& Gulcicek~(\cite{doering89b}), circles.
{\bf (d)} Doering \& Yang~(\cite{doering01}), circles; Johnson~et~al~(\cite{johnson03a}), small filled squares;  model-based results from experimental electron-impact-induced emission cross sections from Johnson~et~al~(\cite{johnson03b}), open squares; 41-state calculation of Tayal~(\cite{tayal02}), thick dotted line.
{\bf (e)} Gulcicek~et~al.~(\cite{gulcicek88}), circles.
{\bf (f)} Gulcicek~et~al.~(\cite{gulcicek88}), circles.
}
\label{fig:comp}
\end{figure*}

The computed cross sections are plotted in Fig.~\ref{fig:cross}.  Note, that in both Figs.~\ref{fig:cross} and~\ref{fig:comp}, some regions which were affected by pseudo-resonances at energies above the ionisation threshold were either smoothed or interpolated to remove any sharp transient behaviour which is unlikely to be physical.  The results of the 38-state and 19-state calculations are shown, both smoothed in the same manner.  Comparison of these results, along with the 33-state calculation, allows pseudo-resonances to be identified as features which change between the calculations thus are very likely to be spurious (due to incompleteness of the model), while features that remain unchanged are most likely real. In general, we may note that the near threshold behaviour, most important for the temperatures of interest, are not greatly affected.  However, it is clear that many of the cross sections are probably affected by pseudo-resonances particularly at {\ch total} energies around 15-20~eV.  Note, the feature at around 14~eV in the $^1D$--$^1$S cross section is a real feature, it is a resonance in the $^2$S partial cross section (see e.g.\ Thomas~et~al.~\cite{thomas97}). 

As expected, the singlet-quintet transitions have null cross-sections since in $LS$-coupling total electronic spin must be conserved.  In addition, the $^1S$$\rightarrow$$^3S^o$ transition is forbidden as $S^e$$\leftrightarrow$$S^o$ transitions are forbidden in general (Goddard~et~al.~\cite{goddard71}).

Experimental results at low energy are available for some transitions from the ground state.  We compare our results with experiment in Fig.~\ref{fig:comp}.  
In the case of the $^3P$$\rightarrow$$^3S^o$ transition, we also plot the cross section from the 41-state $R$-matrix with pseudostates calculation tabulated in Johnson~et~al~(\cite{johnson03a}) and described in Tayal~(\cite{tayal02}).  

In general the computed data compare quite well with the experimental data considering the experimental error bars.  For the $^3P$$\rightarrow$$^1D$ transition, the experimental results suggest a peak in the cross section at an impact energy of about 6~eV.  A similar peak is seen in the computed cross section, though only roughly half the magnitude.  Compared to the cross section of Thomas~et~al., the peak is about of the same magnitude but at lower energy, with our cross section not showing the slow rate of increase in the cross section of that of Thomas~et~al., but rather a sharp increase at the threshold.  The cases of the $^3P$$\rightarrow$$^1S$ and $^1D$$\rightarrow$$^1S$ (see Fig.~\ref{fig:cross}) transitions are quite similar, again rather than the slow rate of increase in the cross section just above threshold of Thomas~et~al., we find a sharp increase.  

The cross section for the $^3P$$\rightarrow$$^5S^o$ transition compares well with experiment at low energy, but is significantly larger than the experimental values at energies at and above 20~eV.  The cross sections also compare well with those of Zatsarinny \& Tayal~(\cite{zatsarinny02}), having the same magnitude and form near threshold.  The peak at about 18~eV seems to be the result of pseudo-resonances, as judged from the fact that the peak varied significantly between the 19-, 33- and 38-state calculations (see Fig.~\ref{fig:cross}).  We note the peak is not present in the results of Zatsarinny \& Tayal, which uses the non-orthogonal orbital method which should be more robust against such pseudo-resonances.  Similarly, the $^3P$$\rightarrow$$^5P$ and $^3P$$\rightarrow$$3p ^3P$ results seem to be affected by pseudo-resonances at {\ch total energies} of about 18~eV.  However, near threhold the results are similar to those of Zatsarinny \& Tayal.

The resonance transition $^3P$$\rightarrow$$^3S^o$ is the best studied, with a significant number of experimental and theoretical studies.  As seen in Fig.~\ref{fig:comp}, our result compares well with the experimental results, both in terms of magnitude and form of the cross sections.  The cross section compares well with the theoretical result of Tayal~(\cite{tayal02}) at energies between 15 and 30~eV, which is also plotted.  Our cross sections are around 5--10\% larger, as expected (see discussion in previous section) due to our neglect of a large part of the coupling to the continuum.  Regarding the near threshold behaviour, our results are very similar to those of Zatsarinny \& Tayal~(\cite{zatsarinny02}), the cross sections having very similar magnitude and form, including a sharp resonance at the threshold.

Rate coefficients are computed by integrating a Maxwellian velocity distribution with the cross-sections, and are presented in Table~\ref{tab:ratecoeffs}.  The calculations are done assuming $LS$ coupling, and as such no fine structure is included.  Fine structure of certain levels of interest are often included in non-LTE modelling, and one may wish to somehow redistribute the rate coefficent between fine structure levels.  It is often the practice to set fine structure transition rates to large values such that the fine structure levels have populations according to their statistical weights (e.g. Kiselman~\cite{kiselman93}).  As such, it is irrelevant how the rates are distributed so long as the total rate is conserved.  However, we note that even neglecting relativistic effects in the Hamiltonian, a simple redistribution of the rate coefficients is not possible as a transformation from $LS$ to $J_aj$ coupling (where $J_a$ and $j$ are the initial total angular momenta of the atom and scattered electron respectively) must be done for each partial wave and scattered electron angular momentum (e.g.\ Saraph~\cite{saraph70}).

\begin{table*}
\begin{center}
\caption{Rate coefficients $\langle \sigma v \rangle$, in units of cm$^3$~s$^{-1}$, for electron-impact excitation of neutral oxygen, for selected temperatures in the {\ch range $T=1000$--$50000$~K}.  Note, $a(b) = a \times 10^b$.}
\label{tab:ratecoeffs}
\begin{tabular}{lrrrrrr}
\hline
\hline 
                                                        Initial & \multicolumn{6}{c}{Final state} \\
      state  & 2p$^4$ $^1D  $&2p$^4$ $^1S  $&    3s $^5S^o$&    3s $^3S^o$&    3p $^5P  $&     3p $^3P  $             \\
\hline
 & \multicolumn{6}{c}{\underline{ 1000~K}} \\
2p$^4$ $^3P  $&   $5.72(-21)$&   $5.13(-31)$&   $1.03(-55)$&   $3.40(-57)$&   $6.35(-64)$&    $5.48(-65)$             \\
2p$^4$ $^1D  $&              &   $1.91(-20)$&         $ 0.$&   $3.34(-49)$&         $ 0.$&    $6.34(-56)$             \\
2p$^4$ $^1S  $&              &              &         $ 0.$&         $ 0.$&         $ 0.$&    $8.85(-45)$             \\
    3s $^5S^o$&              &              &              &   $9.98(-10)$&   $1.11(-15)$&    $1.98(-17)$             \\
    3s $^3S^o$&              &              &              &              &   $9.91(-14)$&    $3.44(-15)$             \\
    3p $^5P  $&              &              &              &              &              &    $2.56(-09)$             \\
 & \multicolumn{6}{c}{\underline{ 3000~K}} \\
2p$^4$ $^3P  $&   $2.98(-13)$&   $4.28(-17)$&   $9.33(-25)$&   $3.46(-25)$&   $9.03(-28)$&    $4.89(-28)$             \\
2p$^4$ $^1D  $&              &   $4.17(-13)$&         $ 0.$&   $6.87(-24)$&         $ 0.$&    $9.66(-26)$             \\
2p$^4$ $^1S  $&              &              &         $ 0.$&         $ 0.$&         $ 0.$&    $4.53(-22)$             \\
    3s $^5S^o$&              &              &              &   $2.36(-08)$&   $6.23(-10)$&    $4.43(-11)$             \\
    3s $^3S^o$&              &              &              &              &   $1.65(-09)$&    $9.26(-10)$             \\
    3p $^5P  $&              &              &              &              &              &    $3.51(-08)$             \\
 & \multicolumn{6}{c}{\underline{ 5000~K}} \\
2p$^4$ $^3P  $&   $1.38(-11)$&   $3.07(-14)$&   $1.30(-18)$&   $7.53(-19)$&   $1.53(-20)$&    $1.26(-20)$             \\
2p$^4$ $^1D  $&              &   $1.24(-11)$&         $ 0.$&   $6.11(-19)$&         $ 0.$&    $9.93(-20)$             \\
2p$^4$ $^1S  $&              &              &         $ 0.$&         $ 0.$&         $ 0.$&    $1.41(-17)$             \\
    3s $^5S^o$&              &              &              &   $3.89(-08)$&   $1.02(-08)$&    $8.34(-10)$             \\
    3s $^3S^o$&              &              &              &              &   $1.09(-08)$&    $1.29(-08)$             \\
    3p $^5P  $&              &              &              &              &              &    $5.83(-08)$             \\
 & \multicolumn{6}{c}{\underline{ 8000~K}} \\
2p$^4$ $^3P  $&   $1.24(-10)$&   $1.34(-12)$&   $3.42(-15)$&   $2.66(-15)$&   $1.72(-16)$&    $1.88(-16)$             \\
2p$^4$ $^1D  $&              &   $8.25(-11)$&         $ 0.$&   $3.75(-16)$&         $ 0.$&    $2.63(-16)$             \\
2p$^4$ $^1S  $&              &              &         $ 0.$&         $ 0.$&         $ 0.$&    $7.98(-15)$             \\
    3s $^5S^o$&              &              &              &   $4.57(-08)$&   $4.97(-08)$&    $3.95(-09)$             \\
    3s $^3S^o$&              &              &              &              &   $2.86(-08)$&    $5.65(-08)$             \\
    3p $^5P  $&              &              &              &              &              &    $6.98(-08)$             \\
 & \multicolumn{6}{c}{\underline{12000~K}} \\
2p$^4$ $^3P  $&   $4.31(-10)$&   $1.10(-11)$&   $2.53(-13)$&   $2.47(-13)$&   $2.99(-14)$&    $3.96(-14)$             \\
2p$^4$ $^1D  $&              &   $2.31(-10)$&         $ 0.$&   $1.50(-14)$&         $ 0.$&    $2.18(-14)$             \\
2p$^4$ $^1S  $&              &              &         $ 0.$&         $ 0.$&         $ 0.$&    $3.30(-13)$             \\
    3s $^5S^o$&              &              &              &   $4.48(-08)$&   $1.20(-07)$&    $8.36(-09)$             \\
    3s $^3S^o$&              &              &              &              &   $4.38(-08)$&    $1.26(-07)$             \\
    3p $^5P  $&              &              &              &              &              &    $6.89(-08)$             \\
 & \multicolumn{6}{c}{\underline{20000~K}} \\
2p$^4$ $^3P  $&   $1.18(-09)$&   $5.84(-11)$&   $7.37(-12)$&   $9.56(-12)$&   $1.91(-12)$&    $3.00(-12)$             \\
2p$^4$ $^1D  $&              &   $5.15(-10)$&         $ 0.$&   $3.24(-13)$&         $ 0.$&    $8.02(-13)$             \\
2p$^4$ $^1S  $&              &              &         $ 0.$&         $ 0.$&         $ 0.$&    $6.05(-12)$             \\
    3s $^5S^o$&              &              &              &   $3.74(-08)$&   $2.38(-07)$&    $1.30(-08)$             \\
    3s $^3S^o$&              &              &              &              &   $5.25(-08)$&    $2.31(-07)$             \\
    3p $^5P  $&              &              &              &              &              &    $5.81(-08)$             \\
 & \multicolumn{6}{c}{\underline{50000~K}} \\
2p$^4$ $^3P  $&   $2.71(-09)$&   $2.35(-10)$&   $1.32(-10)$&   $3.04(-10)$&   $8.95(-11)$&    $1.77(-10)$             \\
2p$^4$ $^1D  $&              &   $1.05(-09)$&         $ 0.$&   $5.85(-12)$&         $ 0.$&    $2.57(-11)$             \\
2p$^4$ $^1S  $&              &              &         $ 0.$&         $ 0.$&         $ 0.$&    $6.50(-11)$             \\
    3s $^5S^o$&              &              &              &   $2.03(-08)$&   $4.12(-07)$&    $1.26(-08)$             \\
    3s $^3S^o$&              &              &              &              &   $4.02(-08)$&    $3.62(-07)$             \\
    3p $^5P  $&              &              &              &              &              &    $3.23(-08)$             \\
\hline
\end{tabular}
\end{center}
\end{table*}

It is of interest to compare these results, at least in broad terms, with those that have been used in past non-LTE calculations.  {\ch Below we make comparisons at $T=5000$~K, which is representative of the typical temperatures of interest in cool stellar atmospheres.}  First we make a comparison with the predictions of approximate yet general and simple formulae used by Allende Prieto~et~al.~(\cite{allende03}); this paper may be consulted for the formulae.  For allowed radiative transitions they used a formula due to Mihalas~(\cite{mihalas72}), which has been derived from the approximate formula of van Regemorter~(\cite{vanregemorter62}) adopting an effective Gaunt coefficient of 0.25.  For forbidden transitions an effective collision strength of $\Upsilon_{ij}=0.05$ (for excitation) has been arbitrarily adopted.  The collision rate coefficient is computed from the effective collision strength via what Allende Prieto~et~al.\ have referred to as the Eissner-Seaton formula (Eissner \& Seaton~\cite{eissner74}). We calculated rate coefficients at 5000~K, and these along with the ratio of these values to our values are presented in Table~\ref{tab:allende}.  We see for allowed transitions, the ratios range from about 0.6 to 3.0.  As mentioned, in $LS$-coupling transitions between singlet and quintet states, and the $^1S \rightarrow ^3S^o$ transition, are not collisionally allowed and thus we cannot make a comparison here.  For the remaining forbidden radiative transitions, the ratios range between roughly 0.003 and 60 indicating much poorer agreement in general.  Based on the evidence here, it appears that the approximate formula based on van Regemorter's formula gives at best order of magnitude estimates, while the simple approximation based on an arbitrarily chosen constant $\Upsilon_{ij}$, is considerably less reliable.  

We also compared with the data in the model atom prepared by Carlsson \& Judge~(\cite{carlsson93}) at similar temperatures.  The first notable aspect of the comparison is that for several transitions no collisional coupling was included due to lack of data.  For the transitions $^3P$$\rightarrow$$^1D$, $^1S$ and $^1D$$\rightarrow$$^1S$ they used data from a compilation by Mendoza~(\cite{mendoza81}), which in fact originate from Berrington \& Burke~(\cite{berrington81}).  At 5000~K, the corresponding excitation rate coefficients are $1.82\times10^{-11}$, $1.27\times10^{-14}$ and $1.04\times10^{-11}$~cm~s$^{-1}$ respectively. The ratios with respect to our values are roughly 1.3, 0.4 and 0.8 respectively.  For the $^3P$$\rightarrow$$^5S^o$, $^3S^o$ transitions, they calculated collision rates from the cross sections of Rountree~(\cite{rountree77}).  The cross sections of Rountree are significantly larger at the threshold than modern calculations, including our results, resulting in their rates being a factor of about 2.5 greater than ours.  For the $^5S^o$$\rightarrow$$^5P$ and $^3S^o$$\rightarrow$$^3P$ transitions, Carlsson \& Judge computed rates using the impact parameter method of Seaton~(\cite{seaton62}).  The rates are about a factor of five and two greater than ours, respectively.  

\begin{table*}
\begin{center}
\caption{Rate coefficients $\langle \sigma v \rangle$ in units of cm$^3$~s$^{-1}$ for electron-impact excitation of neutral oxygen for $T=5000$~K calculated using approximate formulae following Allende~Prieto~et~al.~(\cite{allende03}), and the ratios of these rate coefficients to our rate coefficients in Table~\ref{tab:ratecoeffs} ($\langle \sigma v \rangle_\mathrm{approx} / \langle \sigma v \rangle_\mathrm{this\,work}$).  Note, $a(b) = a \times 10^b$.}
\label{tab:allende}
\begin{tabular}{lrrrrrr}
\hline
\hline
                                                        Initial & \multicolumn{6}{c}{Final state} \\
      state  & 2p$^4$ $^1D  $&2p$^4$ $^1S  $&    3s $^5S^o$&    3s $^3S^o$&    3p $^5P  $&     3p $^3P  $             \\
\hline
 & \multicolumn{6}{c}{\underline{$\langle \sigma v \rangle$ from approximate formulae}}\\
2p$^4$ $^3P  $&   $7.04(-12)$&   $4.04(-14)$&   $4.06(-19)$&   $4.91(-19)$&   $1.00(-20)$&    $5.62(-21)$             \\
2p$^4$ $^1D  $&              &   $7.01(-12)$&   $7.04(-17)$&   $2.94(-17)$&   $1.74(-18)$&    $9.75(-19)$             \\
2p$^4$ $^1S  $&              &              &   $6.13(-14)$&   $2.56(-14)$&   $1.51(-15)$&    $8.50(-16)$             \\
    3s $^5S^o$&              &              &              &   $5.11(-10)$&   $2.18(-08)$&    $1.69(-11)$             \\
    3s $^3S^o$&              &              &              &              &   $1.20(-10)$&    $3.79(-08)$             \\
    3p $^5P  $&              &              &              &              &              &    $2.29(-10)$             \\
 &&&&&&\\
 & \multicolumn{6}{c}{\underline{$\langle \sigma v \rangle_\mathrm{approx} / \langle \sigma v \rangle_\mathrm{this\,work}$}}\\
2p$^4$ $^3P  $&   $5.10(-01)$&   $1.28(+00)$&   $3.10(-01)$&   $6.48(-01)$&   $6.48(-01)$&    $4.43(-01)$             \\
2p$^4$ $^1D  $&              &   $5.53(-01)$&             -&   $4.79(+01)$&             -&    $9.77(+00)$             \\
2p$^4$ $^1S  $&              &              &             -&             -&             -&    $5.98(+01)$             \\
    3s $^5S^o$&              &              &              &   $1.31(-02)$&   $2.14(+00)$&    $2.02(-02)$             \\
    3s $^3S^o$&              &              &              &              &   $1.09(-02)$&    $2.94(+00)$             \\
    3p $^5P  $&              &              &              &              &              &    $3.92(-03)$             \\
\hline
\end{tabular}
\end{center}
\end{table*}

{\ch Zatsarinny \& Tayal~(\cite{zatsarinny03}) have published effective collision strengths based on calculations using the $B$-spline $R$-matrix approach with non-orthoganal orbitals.  They provide data for transitions between the $^3P$, $^1D$ and $^1S$ states of the ground configuration and from these states to 23 other excited states.  We compare our results with theirs at 5000~K in Table~\ref{tab:zt}.  The data are in good agreement, agreeing within a factor of 3 at worst, often much better, the mean ratio $\langle \sigma v \rangle_\mathrm{ZT} / \langle \sigma v \rangle_\mathrm{this\,work}$ being 1.23 with a standard deviation of 0.67.}

Estimating the errors in the computed rate coefficients is difficult.  The generally good agreement of our cross section results with experiment, particularly at low energy, would indicate that our rate coefficients are of quite acceptable accuracy.   Errors due to neglect of coupling to the continuum and effects of pseudo-resonances at energies above the ionisation threshold are expected to be small as these energies hardly contribute to the rate coefficient for the temperatures of interest.  {\ch Perhaps the best estimate of the error is obtained from the scatter among different $R$-matrix calculations where available. From the comparison with the rates of Zatsarinny \& Tayal~(\cite{zatsarinny03}) and the rates of Berrington \& Burke~(\cite{berrington81}) we would estimate that the rate coefficients have errors of the order of 70\% at the temperatures of interest.}

\begin{table*}
\begin{center}
\caption{{\ch Rate coefficients $\langle \sigma v \rangle$ in units of cm$^3$~s$^{-1}$ for electron-impact excitation of neutral oxygen for $T=5000$~K from Zatsarinny \& Tayal~(\cite{zatsarinny03}) for transitions from the ground configuration, and the ratios of these rate coefficients to our rate coefficients in Table~\ref{tab:ratecoeffs} ($\langle \sigma v \rangle_\mathrm{ZT} / \langle \sigma v \rangle_\mathrm{this\,work}$).  Note, $a(b) = a \times 10^b$.}}
\label{tab:zt}
\begin{tabular}{lrrrrrr}
\hline
\hline
                                                        Initial & \multicolumn{6}{c}{Final state} \\
      state  & 2p$^4$ $^1D  $&2p$^4$ $^1S  $&    3s $^5S^o$&    3s $^3S^o$&    3p $^5P  $&     3p $^3P  $             \\
\hline
 & \multicolumn{6}{c}{\underline{$\langle \sigma v \rangle$ from Zatsarinny \& Tayal}}\\
2p$^4$ $^3P  $&   $2.05(-11)$&   $1.16(-14)$&   $1.94(-18)$&   $1.15(-18)$&   $3.42(-20)$&    $2.86(-20)$             \\
2p$^4$ $^1D  $&              &   $7.79(-12)$&         $ 0.$&   $3.45(-19)$&         $ 0.$&    $7.29(-20)$             \\
2p$^4$ $^1S  $&              &              &         $ 0.$&         $ 0.$&         $ 0.$&    $1.63(-17)$             \\
 &&&&&&\\
 & \multicolumn{6}{c}{\underline{$\langle \sigma v \rangle_\mathrm{ZT} / \langle \sigma v \rangle_\mathrm{this\,work}$}}\\
2p$^4$ $^3P  $&   $1.49(+00)$&   $3.67(-01)$&   $1.48(+00)$&   $1.52(+00)$&   $2.21(+00)$&    $2.25(+00)$             \\
2p$^4$ $^1D  $&              &   $6.15(-01)$&             -&   $5.62(-01)$&             -&    $7.31(-01)$             \\
2p$^4$ $^1S  $&              &              &             -&             -&             -&    $1.14(+00)$             \\
\hline
\end{tabular}
\end{center}
\end{table*}

It is difficult to predict the impact that these new data will have on the results of non-LTE calculations in stellar atmospheres.  The impact on abundance analysis of solar-type stars, including metal-poor stars, is presently being investigated by Fabbian~et~al.~(in preparation).  It would also be of interest to investigate the impact on predicted solar centre-to-limb behaviour of the oxygen triplet lines (see e.g.\ Allende Prieto~et~al.~\cite{allende04}).   

{\ch We have presented estimates of cross sections and rate coefficients for all transitions between the seven lowest $LS$ states of neutral oxygen.  It should be noted that the calculations produce data for transitions between all target states listed in Table~\ref{tab:states}.  We restricted the data presented here to the lowest seven states for two reasons.  First, the quality of the results for transitions involving higher states is certainly lower.  In particular, the effect of pseudo-resonances on the rate coefficients would be expected to be larger, since the thresholds are nearer the ionisation energy.  Secondly, the transition rates among the low-lying states are expected to be most important for the non-LTE problem of interest, in particular the modelling of the 777~nm triplet 3s $^5S^o \rightarrow$ 3p $^5P$.  However, the computed rates for the transitions involving the higher states should give at least order of magnitude estimates and may be obtained on request from the author.}  

More calculations and experimental data for electron-impact excitation of oxygen would be valuable, particularly from excited states where few or no other comparisons are available.  {\ch In particular calculations for transitions between excited states using the $B$-spline $R$-matrix method with non-orthogonal orbitals, with more careful consideration of short range correlation and long range polarisation effects would be important.}  Calculations for excitation by hydrogen collisions are also needed.  Recent work by Krems~et~al.~(\cite{krems06}) calculated rates for a single transition, but again {\ch as for electrons}, for stellar atmospheres applications, data are {\ch needed} for more low-lying states.

\begin{acknowledgements}

I thank Martin Asplund and Damian Fabbian for encouragement and useful discussions.  I acknowledge the support of the Swedish Research Council. 

\end{acknowledgements}

\end{document}